%% file: corrloi.tex
\newif\ifcomment
\newif\ifarxiv
\arxivtrue
\newif\ifprint
\printtrue

\ifarxiv
\documentclass[a4paper]{jpconf}
\else
\documentclass[a4paper]{jpconf}
\fi
\usepackage[latin1]{inputenc} 
\usepackage{floatflt}
\usepackage{tabularx}
\usepackage[square,numbers,sort&compress,nonamebreak]{natbib}
\usepackage{hypernat}
\usepackage{doc}
\usepackage{url}
\usepackage{ae}
\usepackage{array}
\usepackage{times}
\usepackage{color}

\newif\ifpdf
\ifx\pdfoutput\undefined
 \pdffalse
\else
 \pdftrue 
\fi

\ifpdf
\usepackage[pdftex]{graphicx}

\ifprint
 \usepackage[pdftex,colorlinks,a4paper,
             linkcolor=blue,citecolor=blue,anchorcolor=blue,
             pagecolor=blue,urlcolor=blue]{hyperref}
\else
 \usepackage[pdftex,backref,pagebackref,colorlinks,
             linkcolor=blue,citecolor=blue,anchorcolor=blue,
             pagecolor=blue,urlcolor=red,a4paper]{hyperref}
\fi
\pdfoutput=1        
\pdfcatalog{/PageMode(/UseOutlines)} 
\else 
\usepackage[dvips]{graphicx}
\usepackage{rotating}

\ifprint
 \usepackage[dvips,colorlinks,a4paper,
             linkcolor=blue,citecolor=blue,anchorcolor=blue,
             pagecolor=blue,urlcolor=blue]{hyperref}

\else
 \usepackage[ps2pdf,backref,pagebackref,colorlinks,
            linkcolor=blue,citecolor=blue,anchorcolor=blue,
            pagecolor=blue,urlcolor=red,a4paper]{hyperref}
\fi
\hypersetup{
  pdfauthor   = {Constantinos A. Loizides},
  pdftitle    = {Jet correlation measurement in heavy-ion collisions: from RHIC to LHC},
  pdfcreator  = {},
  pdfproducer = {},
  pdfsubject  = {},
  pdfkeywords = {}
 }
\fi

\graphicspath{{./pics/}}
\input{commands}

\ifprint
 
\fi

\begin{document}

\title{Jet correlation measurement in heavy-ion collisions:\\ from RHIC to LHC} 
\author{Constantinos A.~Loizides}
\address{Massachusetts Institute of Technology, 77 Mass Ave, Cambridge, 02139, MA, USA}
\ead{loizides@mit.edu}

\begin{abstract}
We attempt to deduce simple options of `jet quenching' phenomena in heavy-ion collisions
at $\snn=5.5~\tev$ at the LHC from the present knowledge of leading-hadron suppression 
at RHIC energies. In light of the nuclear modification factor for leading particles 
we introduce the nuclear modification factor for jets, $\RAA^{jet}$, and for the 
longitudinal momenta of particles along the jet axis, $\RAA^{p_{\rm L}}$.
\end{abstract}

\section{Introduction}
At \ac{RHIC} energies, hard probes, mainly light quarks and gluons, are 
experimentally accessible in heavy-ion collisions with sufficiently high 
rates. Their production is quantified using inclusive single-particle 
spectra at high momentum ($\pt>2-3~\gev$).
The suppression of these high-momentum (leading) particles in central \AAex\ 
relative to peripheral or \pp\ collisions is one of the major discoveries 
at \ac{RHIC}. 
In \AuAu\ collisions at various \cms\ energies ($\snn=62.4$, $130$ and $200~\gev$) 
the two experiments with high transverse-momentum capabilities, \acs{PHENIX} 
and \acs{STAR}, but also \acs{PHOBOS} and \acs{BRAHMS}, have measured:
\begin{itemize}
\item the suppression of single-particle yields at high $\pt$ 
($\gsim 2-4~\gev$)~\cite{arsene2003,adams2003,adler2003,adler2003b,back2003b};
\item the disappearance, in central collisions, of jet-like 
correlations in the azimuthally-opposite side (away-side) of a 
high-$\pt$ leading particle~\cite{adler2002,adams2004} and, quite recently, 
the reappearance of the particles on the away-side manifested in 
low-momentum hadrons~\cite{wang2004,adams2005};
\end{itemize}
The absence of these effects in \dAu\ collisions at the same 
\cms\ energy ($\snn=200~\gev$)~\cite{adler2003c,adams2003b,arsene2003,back2003c}
confirms that in (central) \AAex\ collisions final-state (as opposed to initial-state)
effects modify the measured particle spectra.

The experimental observations have been explained in terms of various quenching 
models, where the energetic partons produced in the initial hard scattering `lose' 
energy as a consequence of the interaction with the dense, partonic matter. 
Most models implement parton energy loss according to medium-induced 
gluon radiation of a hard parton traversing dense partonic matter of finite
size~\cite{wang1998,wang1998b,wang2002,wang2003,vitev2002b,adil2004,vitev2004,
dainese2004,eskola2004}. 
Also hadronic interactions~\cite{gallmeister2004,hardtke2004} have been investigated 
and partially found to contribute to the observed depletion of the hadron spectra. 

At the \ac{LHC} at 30 times higher \cms\ energy, hard probes will be abundantly 
produced, even at energies of more than one order of magnitude higher than at 
\ac{RHIC}. 
These energetic partons might be identified by their fragmentation 
into hadronic jets of high energy. In contrast with \ac{RHIC}, the initial energy
of $\et>50~\gev$ is high enough to allow the full reconstruction of 
the hadronic jet, even in the underlying heavy-ion environment~\cite{thesisloizides}. 
Therefore, measurements of changes in the properties of 
identified jets in \AAex\ with respect to \pp\ collisions will become possible. 

In the present work we attempt to deduce perspectives for `jet quenching' measurements
at LHC given present data and models of hadron suppression at RHIC. In \sect{leadsec} we
introduce PQM, a model that describes high-$\pt$ suppression at RHIC, and state its prediction
of the nuclear modification factor for LHC. In \sect{leadhadlimits} we argue that the 
surface-emission effect limits the sensitivity of leading particles to the 
density of the medium. In \sect{jetspec} we present a PYTHIA simulation of jet 
quenching at LHC conditions and introduce the nuclear modification factor 
for identified jets, $\RAA^{jet}$, and for the longitudinal momenta of particles 
along the reconstructed jet axis, $\RAA^{p_{\rm L}}$.

\section{Leading-particle spectroscopy at RHIC and LHC conditions}
\label{leadsec}
The effect of the medium on the production of a hard probe
is typically quantified via the nuclear modification factor,
\begin{equation}      
\label{eq:rab}
\RAB(\pt,\eta;\,b)= \frac{1}{\av{\Ncoll(b)}} \times \frac
{\dd^2N^{\rm hard}_{\rm AB}/\dd\pt\,\dd\eta}{\dd^2N^{\rm hard}_{\rm pp}/\dd\pt\,\dd\eta}\;,
\end{equation}
which measures the deviation of the \AAex\ from the superposition 
of independent \NNex\ collisions. In absence of strong nuclear 
initial-state effects it should be unity, if binary collision scaling holds.

\begin{figure}[htb]
\begin{center}
\includegraphics[width=7.5cm, height=5cm]{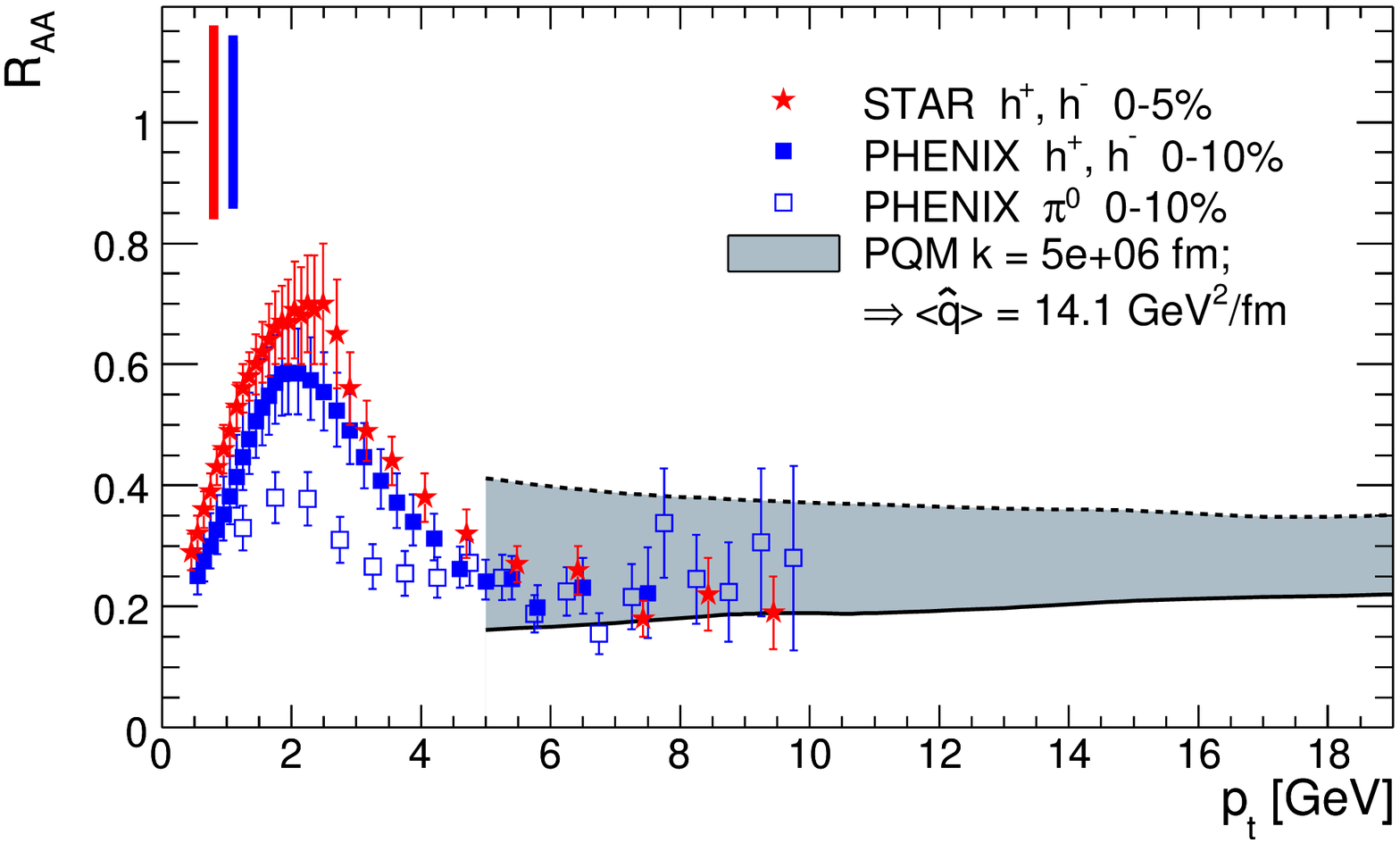}
\hspace{0.5cm}
\includegraphics[width=7.5cm, height=5cm]{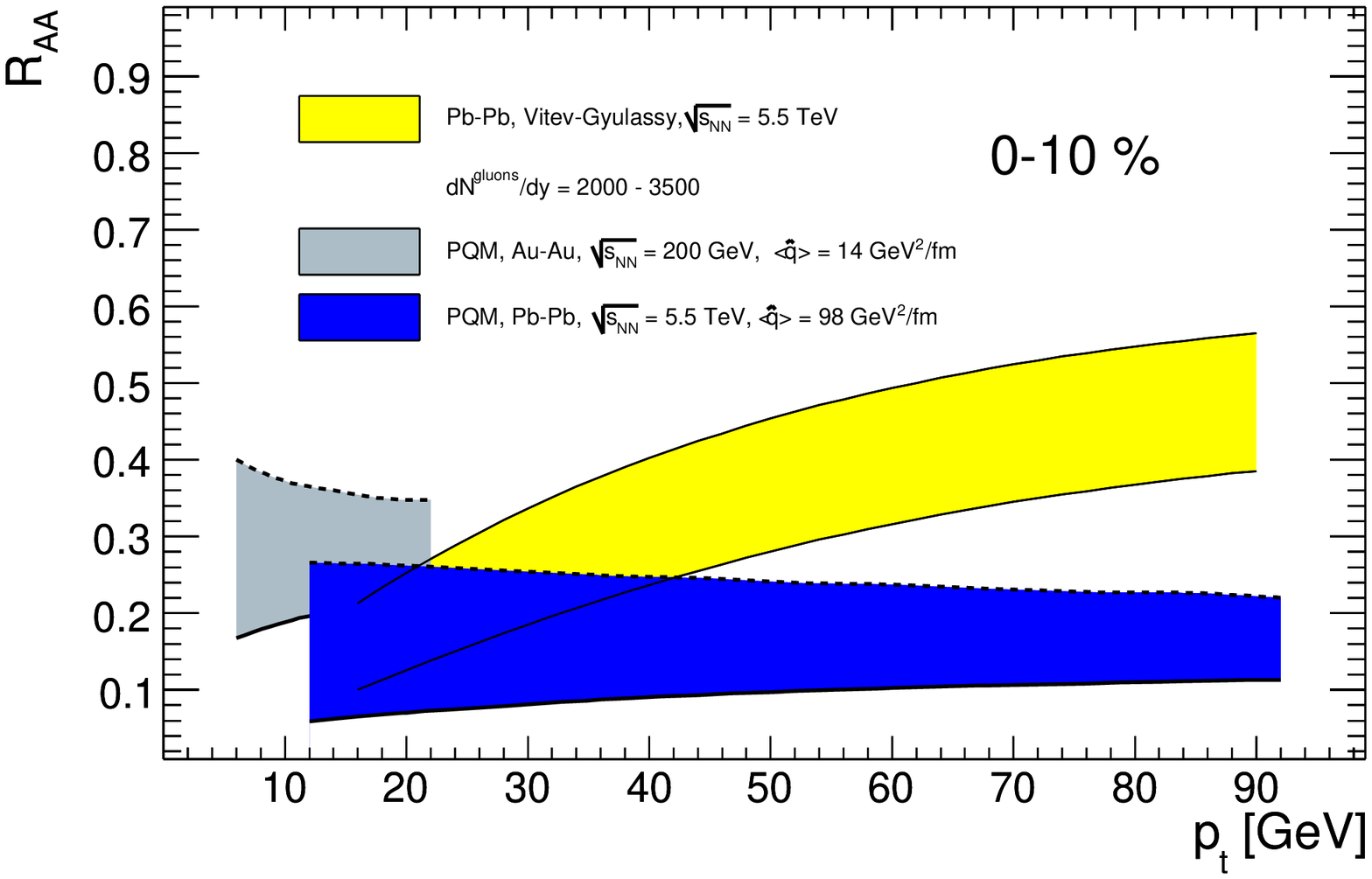}
\caption{\label{fig:raa}
Left: $\RAA(\pt)$ for central \AuAu\ collisions at $\snn=200~\gev$. PHENIX~\cite{adler2003b,adler2003} 
and STAR~\cite{adams2003} data are reported with combined statistical and $\pt$-dependent 
systematic errors (bars on the data points) and $\pt$-independent systematic errors (bars at 
$\RAA=1$). The theoretical calculation is obtained with PQM resulting in an
average transport coefficient of $14~\gev^2/\fm$. The band denotes the systematical
uncertainty of the model.
Right: The PQM prediction for $\RAA(\pt)$ in central \PbPb\ collisions
at $\snn=5.5~\tev$. The LHC prediction presented in 
\Ref{vitev2002b} (Vitev--Gyulassy) and PQM for central \AuAu\ 
collisions at RHIC, $\snn=200~\gev$, are shown for comparison.} 
\end{center}
\end{figure}

\Fig{fig:raa} (left) shows $\RAA(\pt)$ for central \AuAu\ collisions at $\snn=200~\gev$ 
as published by PHENIX~\cite{adler2003b,adler2003} and STAR~\cite{adams2003} together with 
a \ac{pQCD} calculation of parton energy loss obtained with our 
Monte Carlo program PQM (Parton Quenching Model)~\cite{dainese2004}. 

PQM combines a recent calculation of the \ac{BDMPS-Z-SW} parton energy 
loss~\cite{salgado2003} with a realistic description of the collision geometry at 
mid-rapidity. Our approach allows to calculate the transverse momentum and centrality 
dependence of single- and di-hadron correlation suppression, as well as 
the `energy-loss induced' azimuthal anisotropy of particle production in 
non-central collisions. The model has one single parameter, $k$, that sets the 
scale of the density of the medium. Once the parameter is fixed, \eg\ on the basis 
of the nuclear modification data at $\snn=200~\gev$, we scale it to different energies 
and collision systems assuming its proportionality to the expected volume-density of 
gluons. The application of our model is restricted  to the high-$\pt$ region, 
above $4-5~\gev$ at RHIC and above $10~\gev$ at LHC 
energies, since we do not include initial-state effects.

The leading-particle suppression in \AAex\ collisions is obtained by evaluating 
\begin{equation}
\label{eq:medium}
\begin{array}{ll}
\displaystyle{\left.\frac{\dd^2\sigma_{\rm quenched}^h}{\dd\pt\dd \eta}\right|_{\eta=0} =} &
\displaystyle{\sum_{a,b,j=\rm q,\overline q,g}}
\displaystyle{\int}\dd x_a\,\dd x_b\,\dd\Delta E_j\,\dd z_j\, 
f_a(x_a)\,f_b(x_b)\,
\displaystyle{\left.\frac{\dd^2\hat{\sigma}^{ab\to jX}} 
{\dd p^{\rm init}_{{\rm t},j}\dd \eta_j}\right|_{\eta_j=0}}\times \\
&\displaystyle{\times\,
\delta\left(p^{\rm init}_{{\rm t},j}-(p_{{\rm t},j}+\Delta E_j)\right)\,
P(\Delta E_j;R_j,\omega_{{\rm c},j})\,
\frac{D_{h/j}(z_j)}{z_j^2}\,},
\end{array}
\end{equation}
in a Monte Carlo approach. \Eq{eq:medium} describes the production of high-$\pt$ 
hadrons within the perturbative QCD collinear factorization framework,
at mid-rapidity, $\eta=0$, including medium-induced parton energy loss 
in the \ac{BDMPS-Z-SW} formalism. $f_{a(b)}$ is the parton distribution function 
for a parton of type $a(b)$ carrying the momentum fraction $x_{a(b)}$, 
$\hat{\sigma}^{ab\to jX}$ is the
partonic hard-scattering cross sections and $D_{h/j}(z_j)$ is the 
fragmentation function, i.e. the probability distribution for the parton $j$ 
to fragment into a hadron $h$ with transverse momentum 
$\pt=z_j\,p_{{\rm t},j}$. 
The modification with respect to the \pp\ (vacuum) case is given by 
the energy-loss probability, $P(\Delta E_j;R_j,\omega_{{\rm c},j})$, 
for the parton $j$. Its two input parameters, the kinematical constraint
and the characteristic scale of the radiated gluons, depend on the
in-medium path length $L$ of the parton and on the BDMPS transport 
coefficient of the medium, $\hat{q}$. The latter is defined as the average 
medium-induced transverse momentum squared transferred to the 
parton per unit path length. 
(See \Ref{dainese2004} on how we calculate the two parameters $R$ and 
$\omegac$ for a given parton and realistic geometry.)
In the original calculation~\cite{salgado2003}, the energy-loss 
probability (quenching weight) is independent of the energy of the original
parton allowing a parton with finite energy to radiate more than its energy. 
To account for finite parton energies we have introduced two different ways
of constraining the weights, the {\it non-reweighted} and {\it reweighted} 
case. In the first case we constrain the loss to the energy of the parton, 
whenever the radiated energy is determined to be larger than that. In the 
second case we require by truncation of the distribution that the energy loss 
is less than the energy of the parton. 
The resulting energy loss is larger in the {\it non-reweighted} case, 
because partons `thermalize' with a probability
$\int_E^\infty\dd\epsilon\,P(\epsilon)$. It is argued~\cite{salgado2003,eskola2004} 
that the difference in the values of the observables for the two approaches 
illustrates the theoretical uncertainties of the model. 

Coming back to \fig{fig:raa} (left) we note that for the chosen value of $k$
the calculation reproduces $\RAA(\pt)$ for central \AuAu\ collisions at 
$\snn=200~\gev$. In \Ref{dainese2004} we consistently compare model predictions
to various high-$\pt$ observables at RHIC energies. We also show that we need 
to scale $k$ by a factor of seven to obtain the LHC prediction of $\RAA$ at 
$\snn=5.5~\tev$ shown in \fig{fig:raa} (right).
Our prediction for the LHC is in agreement, both in the numerical value and 
in the $\pt$-dependence, with that obtained in \Ref{eskola2004}, 
while it is quite different from the one calculated in \Ref{vitev2002b}.

\section{Properties of leading-hadron suppression}
\label{leadhadlimits}
\begin{figure}[htb]
\begin{center}
\includegraphics[width=12cm]{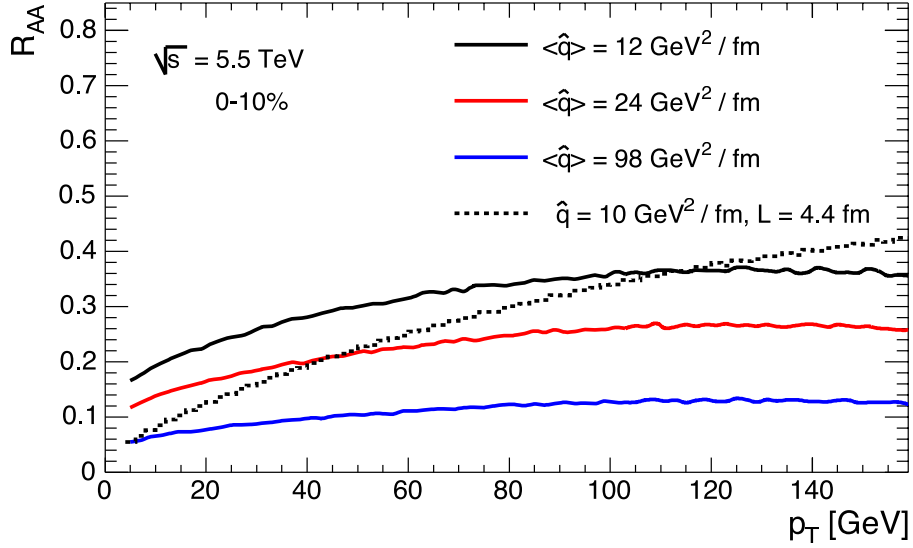}
\end{center}
\vspace{-0.3cm}
\caption{\label{fig:lhcflat}
$\RAA$ as a function of $\pt$ for $0$--$10$\% most central collisions at \acs{LHC} 
energy obtained by \acs{PQM}. The calculations in the parton-by-parton approach (solid lines) 
are compared to a calculation for fixed transport coefficient and length (dashed). 
All graphs are in the non-reweighted case.} 
\end{figure}

At \ac{LHC} energies, the expected $\RAA$ for central \PbPb\ collisions at \ac{LHC} 
as a function of $\pt$ is rather flat, \ie~almost $\pt$-independent, similar
to what is observed at RHIC at $\pt>4~\gev$. 
To study this $\pt$-dependence we show in \fig{fig:lhcflat} $\RAA(\pt)$ in central \PbPb\ at 
\ac{LHC} for $\av{\hat{q}}=12~\gev^2/\fm$ (corresponding to the value found at \ac{RHIC}), 
$\av{\hat{q}}=24~\gev^2/\fm$ and $\av{\hat{q}}=98~\gev^2/\fm$ (the value obtained by the scaling 
from \ac{RHIC} to \ac{LHC}), as well as the result of a calculation with fixed 
$\hat{q}=10~\gev^2/\fm$ and fixed length of $4.4~\fm$. 
Clearly, the fixed-length calculation  shows a stronger $\pt$-dependence than the PQM calculation.

\medskip
As we explain in the following, the flatness of $\RAA$ is a consequence of
\begin{itemize}
\item the steeply falling production cross-section, $\propto \left(\frac{1}{p^{\rm hadron}_{\rm T}}
\right)^{n(\rm \pt)}$, where $n(\pt)$ is rising from about $7$ to $12$ (\ac{RHIC}) and from $6$ 
to $7$ (\ac{LHC}) in the relevant $\pt$ regime; 
\item the emission from the surface, which for large medium densities dominates~\cite{mueller2002}.
\end{itemize}

$\RAA$, \eq{eq:rab}, at mid-rapidity, can be approximated by 
\begin{equation}
\label{eq:raaapprox}
\RAA(\pt) = \left.\int \dd\Delta E\, P(\Delta E,\, \pt+\Delta E) \, \frac{\dd N^{\rm pp}
(\pt+\Delta E)}{\dd\pt} \; \right/ \; \frac{\dd N^{\rm pp}(\pt)}{\dd\pt}\;,
\end{equation}
where, $\dd N^{\rm pp}/\dd\pt$ is the spectrum of hadrons (or partons) in the case of no medium 
(\ie~\pp\ neglecting initial state effects). The suppression computed with \eq{eq:raaapprox} 
is found to give a rather good approximation to the one computed with \ac{PQM} or with the full 
formula, \eq{eq:rab}. 
In the case the production spectrum is (approximately) exponential the $\pt$-dependence cancels 
in the ratio and  we find $\RAA$ to be (approximately) independent of $\pt$. At \ac{RHIC} this is 
the case at about $\pt\ge 30~\gev$. Below that value and at the measured values of
$5$--$12~\gev$, as well as at the \acs{LHC} ($n(\pt)\le7$), the spectrum is given by a power-law.
It was expected~\cite{baier2001} that $\RAA\propto(1+c\,/\sqrt{n(\pt)\,\pt})^{-n(\pt)}$, 
\ie~reaching unity for large $\pt$. 

However, this neglects the fact that for dense media surface emission or, more generally, the 
probability to have no energy-loss, $P(\Delta E=0,\, E)$, plays a significant role, an effect 
which is even more pronounced at low $\pt$ (compared to $\omega_{\rm c}$).
To simplify our argumentation we allow either no loss ($\Delta E=0$) or complete loss ($\Delta E=E$)
in the non-reweighted case, $P(\Delta E,\, E)= p_0 \, \delta(\Delta E) + (1-p_0) \, \delta(\Delta E 
- E)$.~\footnote{For a dense medium the constrained weights at low parton energy indeed do 
have a sharp peak at zero and at maximum possible energy loss, whereas the values in between 
are negligible.}
Inserting the constrained weight into \eq{eq:raaapprox} we obtain
\begin{equation}
\label{eq:raaapproxtoymod}
\RAA(\pt) = p^{*} + (1-p^{*})\, \left. \frac{\dd N^{\rm pp}(2\,\pt)}{\dd\pt} \,
\right/ \, \frac{\dd N^{\rm pp}(\pt)}{\dd\pt}\;.
\end{equation}

\begin{figure}[htb]
\begin{center}
\includegraphics[width=10cm]{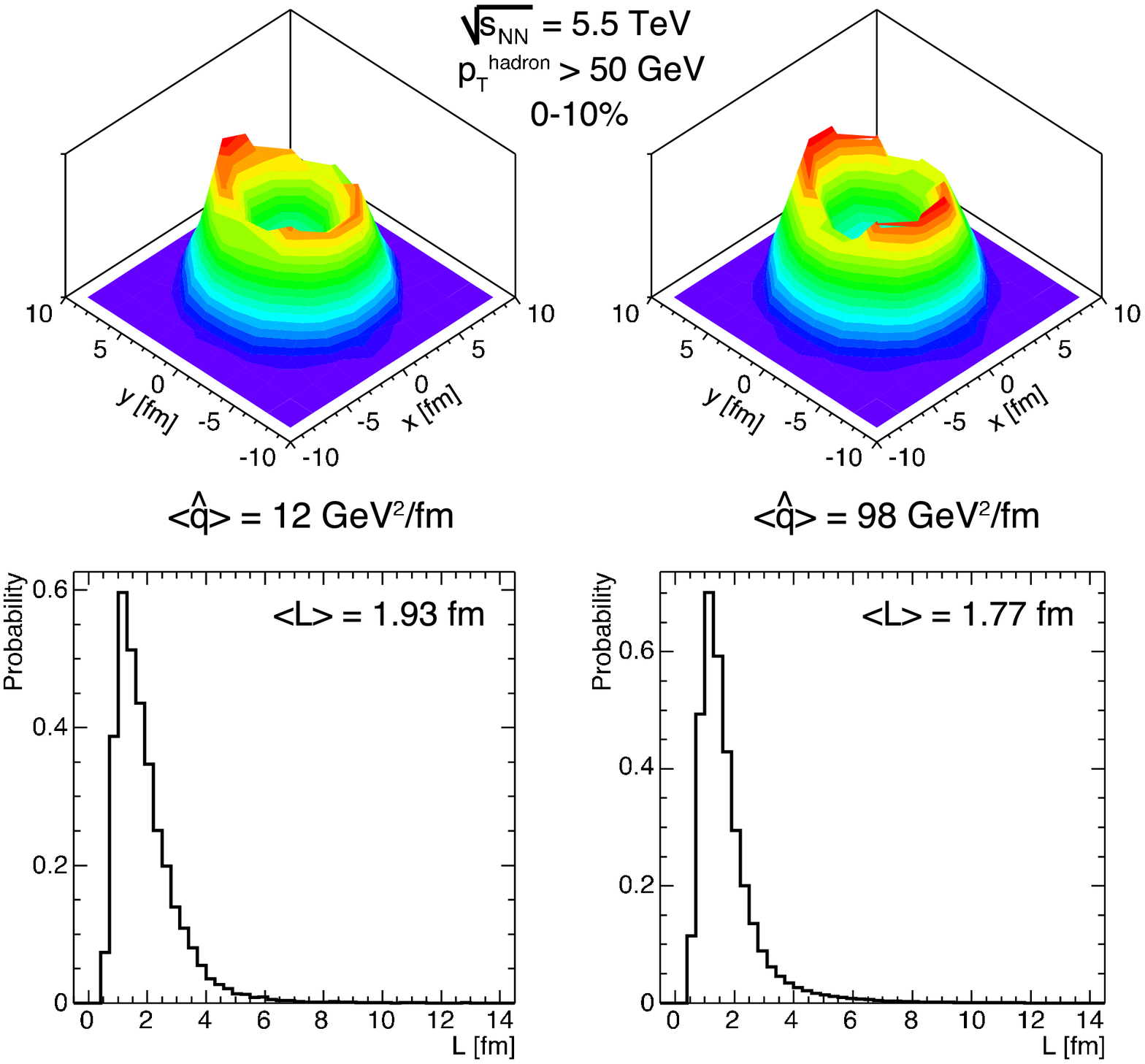}
\end{center}
\caption{\label{fig:surv}
Distributions of parton production points in the transverse plane (upper row)
and in-medium path length (lower row) for partons that escape the medium and produce 
hadrons with $\pt>50~\gev$ in central \PbPb\ collisions at 5.5~$\tev$ for
$\av{\hat{q}}=12~\gev^2/\fm$  (left) and $\av{\hat{q}}=98~\gev^2/\fm$ (right). 
The quantity $\av{L}$ denotes the average of the path-length distribution. The 
calculation is for the non-reweighted case.} 
\end{figure}

\begin{figure}[htb]
\begin{center}
\includegraphics[width=10cm]{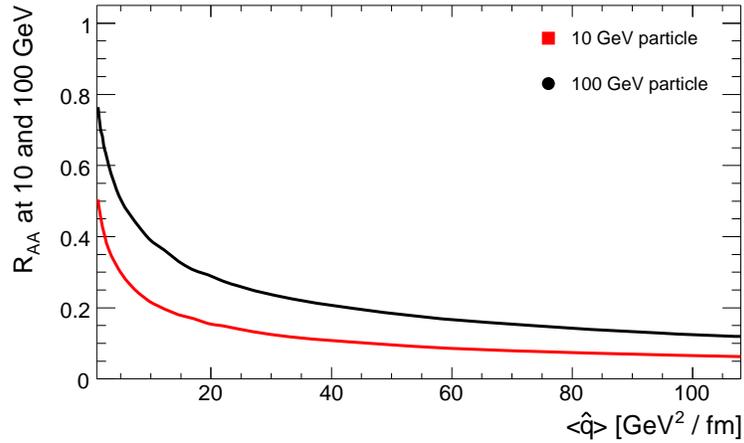}
\end{center}
\caption{\label{fig:raavsq}
$\RAA$ as a function of the average transport coefficient, $\av{\hat{q}}$, for $10$ and 
$100~\gev$ hadrons in $0$--$10$\% central \PbPb\ collisions at $\snn=5.5~\tev$. 
The calculation is for the non-reweighted case.}
\end{figure} 

It is obvious that \eq{eq:raaapproxtoymod} is just a crude approximation, but it demonstrates
that the value of $\RAA$ is dominated by the fraction of hadrons (or partons), which 
escape without losing much of their energy. For the simple power-law production spectrum the 
contribution from higher $\pt$ is suppressed by about $(1+\Delta E/\pt)^{n(\pt)}$.
Taking into account only fixed values of $\hat{q}$ and $L$ the probability $p^{*}$ is given 
by the discrete weight, $p_0$, at $R=0.5\,\hat{q}\,L^3$, amounting to about 0.05(0.002) for 
quarks(gluons). However, for a proper calculation 
one must take into account the right production ratio of quarks-to-gluons. 
For realistic path-length distributions $p^{*}$ is dominated by partons, which are emitted
close to the surface and, thus, enhanced relative to $p_0$ obtained at fixed scale.
It turns out that $p^{*}$ evaluated with PQM at \ac{LHC} central conditions, averaged over path-lengths
and parton types, is independent of $\pt$ in the range shown above, and takes values of about 
$0.14$, $0.1$ and $0.05$ for the scales used in \fig{fig:lhcflat}. 

Surface emission will be present even for very large-momentum hadrons.
\Fig{fig:surv} (top) visualizes the distribution of 
production points ($x_0$,~$y_0$) in the transverse plane for partons, which escape from
the dense overlap region and yield hadrons with $\pt^{\rm hadron}>50~\gev$. The chosen 
values of the medium density correspond to the value found at RHIC and the expectation for
LHC. The corresponding in-medium path-length distributions, \fig{fig:surv} (bottom),
reveal that the average `thickness' of the emission surface is restricted to less than 
$2~\fm$: even hadrons with $\pt>50~\gev$ are emitted dominantly from the 
surface.

The dominance of the surface effect limits the sensitivity of leading particles to the 
density of the medium, mainly for experimentally accessible low-$\pt$ range at \ac{RHIC}. 
This is demonstrated in \fig{fig:raavsq} where we show the dependence of $\RAA$ on the 
average transport coefficient evaluated with \ac{PQM} for \mbox{$0$--$10$\%} most central 
collisions at $\snn=5.5~\tev$ first shown in \Ref{eskola2004}.
For $10~\gev$ hadrons the nuclear modification factor 
is sensitive to average medium densities up to about $15~\gev^2/\fm$, and for $100~\gev$ 
hadrons the sensitive regime might widen to average values of about $30$. 

\section{Jet spectroscopy and modification of jet properties}
\label{jetspec}
At LHC, leading-hadron spectroscopy will be naturally extended by jet spectroscopy.
Simulations show that jets with transverse energies of more than 
$50~\gev$ are identifiable and reconstructible on an event-by-event basis, 
even in most central collisions; however 
with severe limitations in the resolution of the jet energy.
The interaction of the hard partons with the medium created in these collisions
is expected to manifest in the modification of jet properties deviating from known 
fragmentation processes in vacuum. Calculations predict that the additional gluons
radiated by the original parton remain (fragment) inside the jet cone, although 
redistributed in transverse phase space~\cite{wiedemann2000c,salgado2003}. 
The corresponding jet-production cross section is expected to follow binary scaling. 
However, the jet shape is claimed to broaden and the jet multiplicity to soften and 
increase~\cite{salgado2003b}. Ideally, if one reconstructs the hadronic 
energy for outstanding high-energy jets the jet energy may be associated with 
the original energy of the parton. 

In combination with results at lower parton energies from correlation methods
jet measurements at higher energy might complete the picture of medium-induced 
parton-energy-loss phenomena. In the following it is our aim to introduce a simulation
and analysis of jet quenching effects. Details can be found in \Ref{thesisloizides}.
Unquenched jets representing the \pp\ reference are generated using the PYTHIA
generator~\cite{mcpythia2001}. Quenched signal jets are simulated as an afterburner
to \acs{PYTHIA} developed by A.~Morsch. 
It introduces parton energy loss via final-state gluon radiation, in a rather ad-hoc way:
Before the partons originating from the hard $2$-to-$2$ process (and the gluons 
originating from \ac{ISR}) are subject to fragment, they are replaced according 
\[
{\rm parton}_i(E) \rightarrow {\rm parton_i}(E-\Delta E) + n(\Delta E)\,{\rm gluon}(\Delta E/n(\Delta E))
\]
conserving energy and momentum. For each parton, $\Delta E$ is calculated by \ac{PQM} in the 
non-reweighted case and depends on the medium density, parton type and parton production point 
and emission angle in the collision overlap region. 
The number of radiated gluons, $1\le n(\Delta E)\le6$, is determined by the 
condition that each gluon must have less energy than the quenched parton from 
which it was radiated away. First results using this afterburner
at LHC energies have been discussed in~\Ref{morsch2005}.

For the following sample analysis we have prepared events containing jets simulated with 
the modified PYTHIA version for $\av{\hat{q}}=1.2$, $12$ and $24~\gev^2/\fm$, which
are embedded into $0$--$10$\% central \acs{HIJING}~\cite{mchijing1994} events. 
The choice of first value implies only a very little modification of the 
embedded quenched jets compared to embedding of jets with standard \acs{PYTHIA}. 
The second corresponds to the case found to describe the $\RAA$ at \ac{RHIC}, 
whereas the third is a conservative choice about a factor of four smaller than the 
expected value extrapolated from \ac{RHIC} to \ac{LHC}.

The jet finding is based on final particles (no detectors effects) at mid-rapidity 
($-1<\eta<1$). We use the standard \acs{ILCA} cone finder~\cite{blazey2000}. However, 
to cope with the soft heavy-ion background we need restrict the cone radius to $R=0.3$. 
For charged particles we require $\pt\ge2~\gev$. The same settings are used for 
the reference jet measurement in \pp. In central \PbPb\ collisions the mean 
reconstructed jet energy is about 80\% for $50~\gev$ jets, and more than 90\% 
for $100~\gev$ jets; the resolution is about 20\% and 10\%, respectively. 

\subsection{Nuclear modification factor for jets}
\begin{figure}[htb]
\begin{center}
\includegraphics[width=10cm]{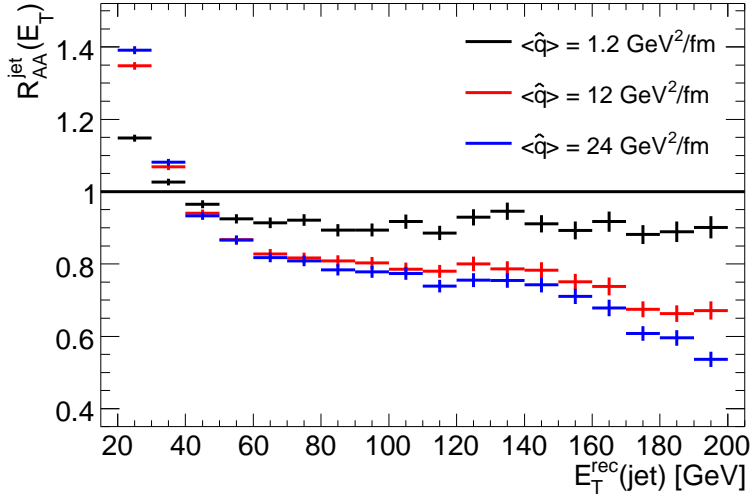}
\end{center}
\caption{\label{raajet}
The nuclear modification factor for jets, $\RAA^{jet}(\et)$, 
as a function of the reconstructed jet energy, $\et^{\rm rec}$, 
for different quenching scenarios in $0$--$10$\% central \PbPb.
The jets in \pp\ and \PbPb\ are identified with the cone finder 
using $R=0.3$ and $\pt>2~\gev$. (Full calorimetry is assumed.)}
\end{figure}

It has been shown~\cite{salgado2003b} that the medium-induced broadening of the jet 
reduces the energy inside $R=0.3$ by $\sim 15$\% and by $\sim7$\% for jets with $\et=50$ and 
$\et=100~\gev$, respectively, and already at $R=0.7$ the effect reduces to about $2$\%. 
However, one must be cautious about these findings, since this prediction has been calculated 
assuming a rather low value of the gluon density at \ac{LHC} conditions. 
We have verified (without embedding in HIJING) that for $R=0.7$ more 
than $90$\% of the energy remains in the jet cone. Up to $45$\% of the rise at jet 
energies lower than $50~\gev$ can be attributed to the contribution of the underlying 
heavy-ion background to the jet signal. 
Jet suppression (at fixed value of $R$) may be observed in the suppression
of inclusive single-jet spectra in central \AAex\ relative to peripheral or \pp\ collisions.
Analogous to \eq{eq:rab}, we therefore define the nuclear modification factor for jets, 
$\RAA^{jet}(\et,\eta;\,b)$, according to 
\begin{equation}      
\label{eq:raajet}
\RAA^{jet}(\et,\eta;\,b)= \frac{1}{\av{\Ncoll(b)}} \times \frac
{\dd^2N^{\rm jet}_{\rm AA}/\dd\et\,\dd\eta}{\dd^2N^{\rm jet}_{\rm pp}/\dd\et\,\dd\eta}\;.
\end{equation}
\Eq{eq:raajet} implicitly depends on the (fixed) value of $R$ used to find the jets.
\Fig{raajet} shows $\RAA^{jet}(\et)$ for different quenching scenarios in $0$--$10$\% 
central \PbPb. 

\subsection{Longitudinal momentum modification factor}
The signature of medium-induced gluon radiation should be visible in 
the modification of the jet fragmentation function as measured through 
the longitudinal and transverse momentum distributions of associated 
hadrons within the jet. 
The momenta parallel to the jet axis, 
$p_{\rm L} = p_{\rm hadron} \, \cos(\theta_{\rm jet},\theta_{\rm hadron})$, 
are expected to be reduced~(jet quenching), while the momenta in the transverse 
direction, $j_{\rm T} = p_{\rm hadron} \, \sin(\theta_{\rm jet},\theta_{\rm hadron})$, 
to be increased~(transverse heating). In our simulation for technical reasons we can not 
give an explicit transverse-momentum kick to the radiated gluons. Therefore, we concentrate 
on the changes of the longitudinal fragmentation.

\begin{figure}[htb]
\begin{center}
\includegraphics[width=10cm]{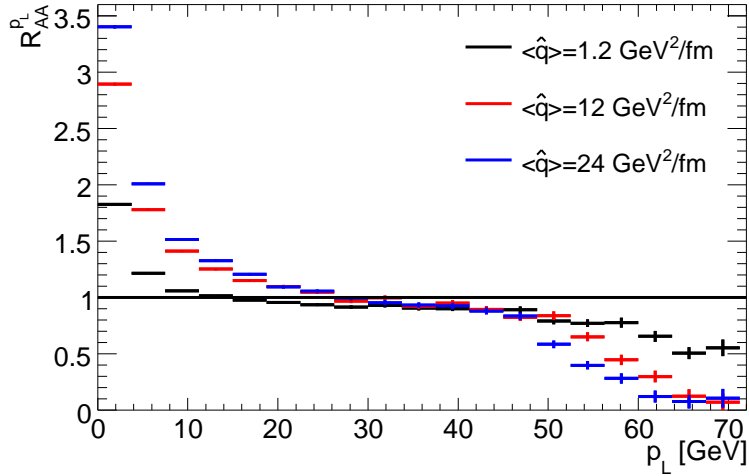}
\end{center}
\caption{\label{fig:raapl}
Longitudinal momentum modification factor, $\RAA^{p_{\rm L}}$, 
for jets with $\et^{\rm rec}>80~\gev$ and different quenching 
scenarios in $0$--$10$\% central \PbPb.
The jets in \pp\ and \PbPb\ are identified with the cone finder 
using $R=0.3$ and $\pt>2~\gev$. (Full calorimetry is assumed.)}
\end{figure}

We define the longitudinal momentum modification factor according to
\begin{equation}
\label{eq:raapl}
\RAA^{p_{\rm L}}(p_{\rm L}) = \frac{N^{\rm jets}_{\rm pp}}{N^{\rm jets}_{\rm AA}}\,
\frac{\dd N_{\rm AA}/\dd p_{\rm L}}{\dd N_{\rm pp}/\dd p_{\rm L}}
\end{equation}
where for proper normalization $N^{\rm jets}_{\rm AA}$ and $N^{\rm jets}_{\rm pp}$ 
account the total number of jets that were found in \AaAa\ and \pp, respectively.
\Eq{eq:raapl} implicitly depends on the transverse jet-energy range of jets
taking into account in the distributions. \Fig{fig:raapl} shows
$\RAA(\rm p_{\rm L})$ for jets with reconstructed energies of
$\et^{\rm rec}>80~\gev$ in $0$--$10$\% central \PbPb\ compared 
for the different quenching scenarios.

The expected behavior is clearly visible: higher medium density leads to 
stronger suppression for large longtitudinal momenta and enhancing of 
smaller momenta. At low $p_{\rm L}$ the effect becomes becomes most apparent.
At these low momenta there is an moderate additional contribution from particles which 
belong to the underlying heavy-ion background, which amounts to about 20\% for
$p_{\rm L}<5~\gev$. It might be interesting to study the (average) slope of 
$\RAA(\rm p_{\rm L})$ as a function of the reconstructed jet energy.

\section{Conclusions}
We have presented the PQM model~\cite{dainese2004} which is able to describe
high-$\pt$ suppression effects at RHIC. Its application to LHC conditions
leads to the expectation that $\RAA$ is essentially constant with $\pt$, 
also for very high momenta up to $100~\gev$.
The nuclear modification factor for leading hadrons is largely dominated by surface 
effects. This limits its sensitivity to the density of the created medium. 
The measurement of reconstructed jets above the underlying heavy-ion 
background may allow to probe the medium to deeper extents. 
In a simulation of jet quenching at LHC we have studied two observables:
the nuclear modification factor for jets, $\RAA^{jet}$, and the nuclear
modification factor for the longitudinal momenta of particles along the 
jet axis, $\RAA^{p_{\rm L}}$. The definitions implicitly depend on the
parameters used to find the jets. For jets with $\et\gsim50~\gev$ both 
are suitable measures to quantify deviations from the \pp\ case:
$\RAA^{jet}(\et)$ decreases with increasing $\et$ and $\RAA^{p_{\rm L}}$ 
is enhanced at low $p_{\rm L}$ and suppressed at high $p_{\rm L}$. 
However, for jets with $\et<50~\gev$ suffer from the contamination of the 
jet cone by uncorrelated particles from the underlying background which 
at these energies severely influences measured jet properties.

\ack
The author acknowledges fruitful discussions with A.~Dainese, 
A.~Morsch, G.~Pai\'c and R.~Stock.

\bibliographystyle{unsrtnat}
\bibliography{corrloi}
\end{document}

%% file: commands.tex
\def\cms    {centre-of-mass}

\newcommand {\ac}[1] {#1}
\newcommand {\acs}[1]{#1}

\newcommand {\snn}{\sqrt{s_{\scriptscriptstyle{{\rm NN}}}}}

\newcommand {\av}[1]{\ensuremath{\left< #1 \right>}}

\newcommand {\gsim}{\,{\buildrel > \over {_\sim}}\,}
\newcommand {\text}[1] {$\mathrm{#1}$}
\newcommand {\hrefurl}[1]{\href{#1}{\url{#1}}}

\newcommand {\Ref}[1]{Ref.~\cite{#1}}

\newcommand {\eq}[1]{eq.~(\ref{#1})}
\newcommand {\Eq}[1]{Equation~(\ref{#1})}
\newcommand {\sect}[1]{section~\ref{#1}}

\newcommand {\fig}[1]{fig.~\ref{#1}}
\newcommand {\Fig}[1]{Figure~\ref{#1}}

\newcommand {\mrm}  {\mathrm}
\renewcommand {\pt}   {\ensuremath{p_{\mathrm{T}}}}

\newcommand {\et}   {\ensuremath{E_{\mathrm{T}}}}

\newcommand {\dd}   {\mrm{d}}
\newcommand {\eg}   {e.g.}
\newcommand {\ie}   {i.e.}

\newcommand {\RAA}  {R_{\rm AA}}
\newcommand {\RAB}  {R_{\rm AB}}

\newcommand {\omegac}{\omega_{\rm c}}

\newcommand {\Ncoll}{\ensuremath{N_{\rm coll}}}

\newcommand {\tev}   {\mbox{${\rm TeV}$}}
\newcommand {\gev}   {\mbox{${\rm GeV}$}}

\newcommand {\mom}   {\mbox{\rm GeV$\kern-0.15em /\kern-0.12em c$}}
\newcommand {\gmom}  {\mbox{\rm GeV$\kern-0.15em /\kern-0.12em c$}}
\newcommand {\mass}  {\mbox{\rm GeV$\kern-0.15em /\kern-0.12em c^2$}}
\newcommand {\mmass} {\mbox{\rm MeV$\kern-0.15em /\kern-0.12em c^2$}}
\newcommand {\mmom}  {\mbox{\rm MeV$\kern-0.15em /\kern-0.12em c$}}

\newcommand {\fm}    {\mbox{${\rm fm}$}}

\newcommand {\pp}    {\mbox{pp}}

\newcommand {\PbPb}  {\mbox{Pb--Pb}}
\newcommand {\AuAu}  {\mbox{Au--Au}}
\newcommand {\AaAa}  {\mbox{A--A}}

\newcommand {\dAu}   {\mbox{d--Au}}

\newcommand {\NNex}  {\mbox{nucleon--nucleon}}

\newcommand {\AAex}  {\mbox{nucleus--nucleus}}

\newcommand {\Jpsi}  {\mbox{J\kern-0.05em /\kern-0.05em$\psi$}}

